\documentclass{pasj00}
\usepackage{graphicx}
\begin{document}
\SetRunningHead{Ryu et al. 2012}{X-ray Echo from Sagittarius~C and 
500-year Activity History of Sagittarius~A*}
\Received{2012/09/04}
\Accepted{2012/10/27}

\author{Syukyo Gando \textsc{Ryu},\altaffilmark{1} Masayoshi \textsc{Nobukawa},\altaffilmark{1,2} Shinya \textsc{Nakashima},\altaffilmark{1} Takeshi Go \textsc{Tsuru},\altaffilmark{1}
\\
Katsuji \textsc{Koyama},\altaffilmark{1,3} 
and Hideki \textsc{Uchiyama}\altaffilmark{4}}  

\altaffiltext{1}{ Department of Physics, Graduate School of Science, Kyoto University, Sakyo-ku, Kyoto, 606-8502}
\email{ryu@cr.scphys.kyoto-u.ac.jp}
\altaffiltext{2}{The Hakubi Center for Advanced Research, Kyoto University, Yoshida-Ushinomiya-cho, Sakyo-ku, Kyoto 606-8302}
\altaffiltext{3}{ Department of Earth and Space Science, Graduate School of Science, Osaka University, 1-1 Machikaneyama, Toyonaka, Osaka, 560-0043}
\altaffiltext{4}{ Department of Physics, School of Science, the University of Tokyo, 7-3-1 Hongo, Bunkyo-ku, Tokyo, 113-0033}

\title{X-ray Echo from the Sagittarius~C Complex and \\
500-year Activity History of Sagittarius A*
} 

\maketitle

\begin{abstract}

This paper presents the Suzaku results obtained for the Sagittarius (Sgr) C region using the concept of X-ray reflection nebulae (XRNe) as the echo of past flares from the super massive 
black hole, Sgr~A*. The Sgr~C complex is composed of several molecular clouds proximately located in projected distance.
The X-ray spectra of Sgr~C were analyzed on the basis of a view that XRNe are located inside the Galactic center plasma X-ray emission with an oval distribution around Sgr~A*. 
We found that the XRNe are largely separated in the line-of-sight
position, and are associated with molecular clouds in different velocity ranges 
detected by radio observations.
We also applied the same analysis to the Sgr~B XRNe and completed 
a long-term light curve for Sgr~A* occurring in the past. 
As a new finding, we determined that Sgr~A* was experiencing periods of high luminosity already $\sim$500 years ago, which is longer than the previously reported value.
Our results are consistent with a scenario that Sgr A* was continuously active with sporadic flux variabilities of $L_{\rm X}=1$--$3\times10^{39}$ erg~s$^{-1}$ in the past 50 to 500 years.
The average past luminosity was approximately 4--6 orders of magnitude higher than that presently observed.
In addition, two short-term flares of 5--10 years are found. 
Thus, the past X-ray flare should not be a single short-term flare, 
but can be interpreted as multiple flares superposed on a long-term high state. 

\end{abstract}

\section{Introduction}
In the Galactic center (GC), prominent 6.4 keV line of neutral iron (Fe\emissiontype{I}~K$\alpha$) has been detected from giant molecular clouds (MCs) such as Sagittarius (Sgr) B ($l\sim\timeform{0.7D}$; \cite{koyama96}), Sgr~C ($l\sim\timeform{359.5D}$; \cite{mura01}; \cite{nakajima09}), and Sgr~A ($l\sim\timeform{0.1D}$; \cite{park04, muno07}). 
The spectra show large equivalent width ($EW_{\rm 6.4~keV}\geq$ 1~keV) and strong absorption ($N_{\rm H} \geq 10^{23}$~H~cm$^{-2}$) to the continuum, 
which suggests that the 6.4~keV bright MCs are due to the irradiation and fluorescence by possible external X-rays 
(X-ray reflection nebulae: XRNe) rather than particle irradiation (e.g., \cite{yusef07}).
Subsequently, the XRN scenario has become more conclusive owing to the discovery of 
the short-term (a few years) 6.4~keV variability in the small (a few light-years) regions in Sgr~B (\cite{koyama08}) and Sgr~A (\cite{ponti10}). 
Moreover, in Sgr~B, the same time variability of hard continuum X-rays ($E\geq8$~keV)
has been found to be in correlation with the 6.4~keV line by \citet{ter10} and \citet{nobu11}. 
These results promoted the idea of the external irradiation source being a past flare of the super massive black hole Sgr~A* (e.g., \cite{koyama96}; \cite{sun93}; \cite{mura01}; \cite{ponti10}; \cite{nobu11}). 

\par
In the XRN context, one can derive the X-ray light curve of Sgr~A* for the past several hundred years. 
The flare luminosity and look-back (delay) time depend on the position of the XRN relative to Sgr~A* and the sun (observer). 
However, most of the previous estimates used solely the projected 
distance or indirect information of line-of-sight positions. 
\par
\citet{ryu09} successfully developed an original method of determining the line-of-sight position for the Sgr~B XRNe. 
Their method involves careful spectral analysis of XRNe in combination with the spatial and 
spectral information of the GC plasma X-ray emission (GCPE).
\citet{uchi11} recently performed an intensive investigation on the 6.7 keV (Fe\emissiontype{XXV} K$\alpha$) line profile for the entire GC region ($\sim\timeform{5D}\times\timeform{2D}$); 
thereafter, \citet{uchi12} extended the work to other emission lines of highly ionized sulfur (S), argon (Ar), and calcium (Ca). 
These studies quantitatively constructed the GCPE as a two-temperature plasma and determined its three-dimensional spatial distribution. 
\par
On the basis of these new pictures of the GCPE, we establish the method to measure the three-dimensional positions of the Sgr~C XRNe. The observation and analyses of the Sgr~C data are given in section~2 and 3.  
We derive a face-on view of XRNe in section~4.1 and compare its reliability with that of radio observations in section~4.2.
We then combine these results with the Sgr~B data and results (\cite{ryu09}) to calculate the time delay and the corresponding 
X-ray flare luminosity to reconstruct the long-term ($\sim$500 years) light curve for Sgr~A*, which occurred in the past. 
The derivation method and discussion for the activity history of Sgr~A* are given in section~4.4.
\par
Throughout this paper, we adopt 8.0~kpc (\cite{ghez08}) as the distance between 
the sun and Sgr~A*; thus, \timeform{1.0D} corresponds to 140~pc at the GC. 
The parameter uncertainties are quoted at the 90\% statistical confidence-level (1.64 $\sigma$) range unless noted otherwise.

\section{Observation}

\begin{table*}[!t]
  \caption{Suzaku Observations at the Sgr C Region.}
  \label{obs}
  \begin{center}
\begin{tabular}{ccccc}
\hline \hline             
Target & Obs. ID      & Obs. point (FOV center) & Obs. date   & Effective   \\ 
name      &           & $\alpha $ (J2000) \hspace{3.5mm}  $ \delta$ (J2000) & (yy-mm-dd) & exposure  \\
\hline
   Sgr C & 500018010 & \timeform{17h44m37.30s}  \hspace{3.5mm}  \timeform{-29D28'10.2''} & 2006-02-20 & 106.9 ks  \\
   Sgr C & 505031010 & \timeform{17h44m58.01s}  \hspace{3.5mm}  \timeform{-29D22'51.2''} & 2010-09-25 &  82.6 ks  \\
\hline

 \end{tabular}
  \end{center}
\end{table*}

  \begin{figure*}[!ht]
  \begin{center}
   \FigureFile(160 mm,){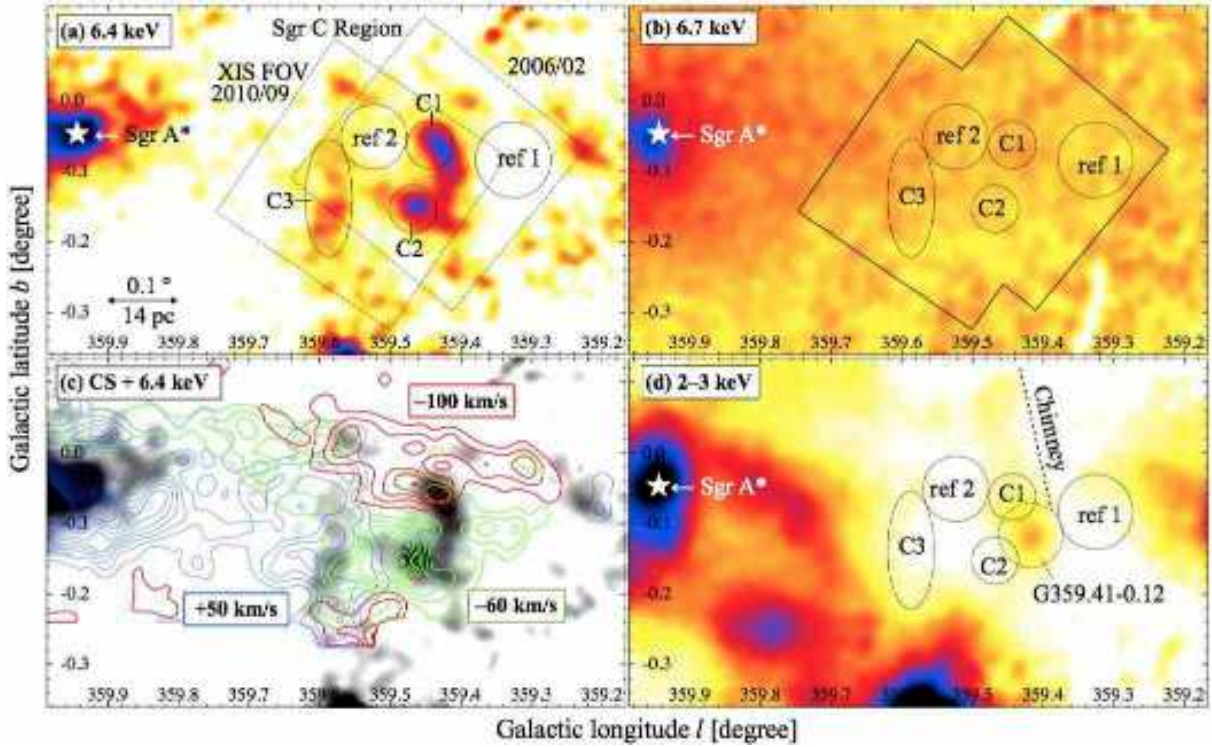}
   \end{center}
  \caption{Added XIS (FI+BI) images near the Sgr C region. 
The images are binned with 12$\times$12 pixels and smoothened with a Gaussian kernel of \timeform{5"}.  
(a) 6.4~keV line (Fe\emissiontype{I}) image obtained by subtracting the continuum component (see text) from the 6.3--6.5~keV band flux. 
The XIS field of views (FOV; \timeform{17.8'}$\times$\timeform{17.8'}) of the 2009 and 2006 observations are represented by the large boxes. 
The regions for spectral analyses (C1--3, ref~1--2) are indicated by solid circles.
(b) Same as panel (a), but for the 6.7 keV line (Fe\emissiontype{XXV}) image.
(c) 6.4~keV line image (gray) with the superposed radio CS ($J$=1$\rightarrow$0) contours (\cite{tsuboi99}) in the velocity range of
 $-100\pm20$ km~s$^{-1}$ (red), $-60 \pm20$ km~s$^{-1}$ (green), and $+50\pm30$ km~s$^{-1}$ (blue). 
The contour level begins at the intensity of 5 K~km~s$^{-1}$ and increases every 5 K~km~s$^{-1}$.
(d) Same as panel (a), but for the 2--3 keV band image (color). 
The region of G359.41$-$0.12 in \citet{tsuru09} is represented by the dashed circle.  
}
\label{sgrc_img}
\end{figure*}

  We have made two deep pointing observations on the Sgr C region separated by $\sim$5 years between February 2006 and September 2010.
These observation are conducted using the X-ray Imaging Spectrometer (XIS) at the focal planes of the X-ray Telescope (XRT) on board the Suzaku satellite. 
The XIS system consists of three sets of front-illuminated (FI)  charge coupled device (CCD) cameras (XIS\,0, 2, and 3) and one set of back-illuminated (BI) CCD camera (XIS~1); each CCD chip contains 1024$\times$1024 pixels (1~pixel = 24~$\mu$m$\;\times \;$24$~\mu$m) for a \timeform{17.8'}$\times$\timeform{17.8'} field of view. 
Two calibration sources of $^{55}$Fe are installed to illuminate two corners of each CCD for absolute gain tuning. 
We operated the XIS in the normal clocking mode with a read-out cycle of 8\,s. The details of Suzaku, XIS, and XRT are given 
in \citet{mitsuda07}, \citet{koyama07a}, and \citet{ser07}, respectively.The observation log of Sgr~C is shown in table~\ref{obs}. Archive data information on the Sgr~B and other relevant GC regions can be found in \citet{ryu09} and \citet{uchi11}.

\section{Data Reduction and Analyses} 

We perform the data reduction and analysis using HEADAS software version 6.12. 
The calibration database used to calculate the response of XIS and the effective area of XRT is the version released\footnote{http://www.astro.isas.ac.jp/suzaku/caldb/} on 2011-02-10.
Since the relative gains and response functions of the FI CCDs are essentially the same, the FI spectra are merged\footnote{The XIS\,2 suddenly became unusable on November 9, 2006, possibly due to a micro-meteoroid impact on the CCD. Therefore, after this epoch, the merged 
FI CCD were XIS\,0 and 3.} in the spectral fitting. 
For the images described in the following sections and spectral analyses, the non-X-ray background (NXB; \cite{tawa08}) is generated using the night-Earth observation data and is subtracted from the raw data. The XRT vignetting effects and the exposure-time differences have been corrected. 

\subsection{Images} 
We construct the XIS images of the Sgr~C region with the 6.4~keV-line (Fe\emissiontype{I}), the 6.7~keV-line (Fe\emissiontype{XXV}), and the 2--3~keV bands (see figure~\ref{sgrc_img}).
The images are generated by adding data from multiple observations together (c.f., table~\ref{obs} and \cite{uchi11}) to cover a wide region near Sgr~C. 
In addition, the FI and BI data are combined to increase the statistics. 
For the 6.4~keV and 6.7~keV line images, the fluxes underlying the continuum are estimated from the 5--6~keV band and subtracted from the 6.3--6.5 and 6.6--6.8 keV bands, respectively. 
Details of the subtraction procedures are explained in \citet{nobu10}.
\par

Figure~\ref{sgrc_img}a displays many bright clumps in the 6.4~keV band near Sgr~C.  
These are spatially associated with giant MCs (see figure~\ref{sgrc_img}c) as observed in the radio CS ($J$=1$\rightarrow$0) line by \citet{tsuboi99}. We designate the brightest three 6.4~keV clumps: C1, C2, and C3 (solid circles; figure~\ref{sgrc_img}a). 
The clumps C1 and C2 were previously identified as XRNe by \citet{nakajima09}, while C3 is a new XRN candidate revealed in the 2010 observation.
\par

Figure~\ref{sgrc_img}b (the 6.7~keV line image) traces the high-temperature plasma (HP) of the GCPE (section 3.2), and shows a more uniform distribution than the 6.4~keV emission shown in figure~\ref{sgrc_img}a.
Figure~\ref{sgrc_img}d shows the 2--3~keV band image including the 2.45~keV line (S\emissiontype{XV}) and displays the distribution of the low-temperature plasma (LP) and individual thermal diffuse sources, mainly supernova remnants candidates, with temperatures of k$T\sim$1~keV.
The faint diffuse sources near the Sgr~C clumps are the supernova remnants G359.41$-$0.12 and its outflow "Chimney" (\cite{tsuru09}). 
To investigate the nature of the GCPE (HP and LP), we selected two regions with relatively weak 6.4~keV emission for spectral analyses, and designate them as ref1 and ref2 (solid circles; figure~\ref{sgrc_img}a, b, and d).

\subsection{Spectral Model} 
\par
\begin{figure}[ht]
  \begin{center}
   \FigureFile(70 mm,){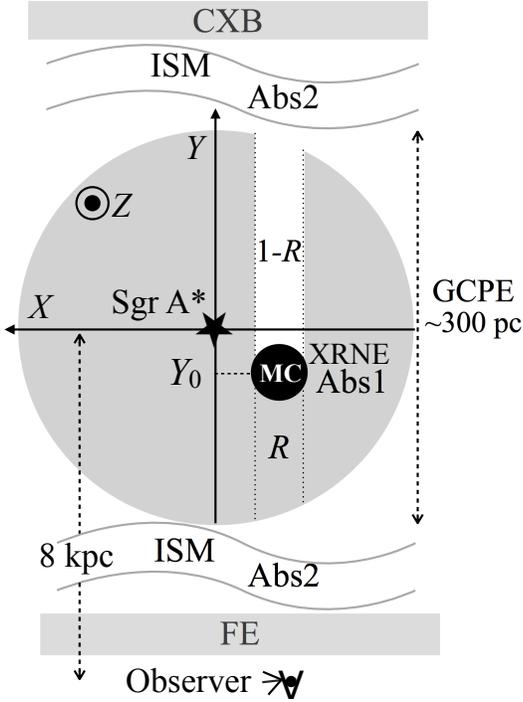}
   \end{center}
  \caption{A schematic face-on view for line-of-sight alignment. 
  Coordinates $X$, $Z$, and $Y$ correspond to directions of $l$, $b$, and 
line of sight, with respect to Sgr~A*. }
\label{los}
\end{figure}

The X-ray spectra of the GC region have been extensively studied by \citet{ryu09} and \citet{uchi12}. 
The spectra can be reproduced by the superposition of four components: the GCPE, 
the XRN emissions (XRNE), the foreground emission (FE), and  the cosmic X-ray background (CXB). 
These components have the following properties.
The GCPE has two-temperature plasmas with k$T$$\sim$1~keV and k$T$$\sim$7~keV for LP and HP, respectively. 
The XRNE is expressed as neutral ion lines (K$\alpha$ and K$\beta$) associated with a power-law component. Since the XRNE is formed by Thomson scattering of a continuum component and fluorescence of iron atoms in the MCs, 
it shows a large absorption (Abs1) of $N_{\rm H}\sim10^{23}$~H~cm$^{-2}$ (e.g., \cite{mura01}), which is approximately equal to the $N_{\rm H}$ through the MC.  
The GCPE including the HP and LP is extended in the GC region, and the MC is inside this region. Thus, a fraction ($R$) of the X-ray from the GCPE is not absorbed by the MC while the other fraction (1$-R$) is absorbed by the MC (Abs1). 
Both GCPE and XRNE exhibit an common interstellar absorption (Abs2) of $N_{\rm H}\sim$ 6 $\times10^{22}$~H~cm$^{-2}$ between the GC and the sun (\cite{sakano02, ryu09}).
The FE is approximated to a thermal plasma of k$T$$\sim$1~keV with small absorption.  
The CXB (e.g., \cite{kushino02}) has the absorption of Abs2 by twice (interstellar absorption in front of and behind the GC), in addition to Abs1 (MC).  
Thus, the spectral model can be expressed by equation~(1).  
A schematic picture of this model is shown in figure~\ref{los}.
{\small
\begin{eqnarray}
f({\it E})\;{\rm = [Abs1\times Abs2 \times (1-{\it R}) + Abs2\times {\it R}]\times (HP+LP) } \hspace{5mm} \nonumber \\
{\rm  + [Abs1 \times Abs2] \times XRNE }  \hspace{40mm} \nonumber \\
{\rm  + [Abs1 \times Abs2 \times Abs2] \times CXB  + FE } \hspace{23.5mm} 
\end{eqnarray}
}
\par
\begin{figure}[!ht]
  \begin{center}
   \FigureFile(70 mm,){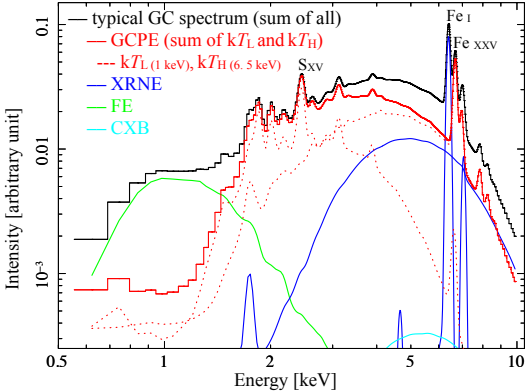}
   \end{center}
  \caption{Simulated XIS/FI spectrum with typical GC parameters reported by \citet{ryu09}. Model components of GCPE 
(k$T_{\rm LP}=$1~keV, k$T_{\rm HP}=$6.5~keV, $\alpha$=0.27, $R$=0.5), XRNE ($EW_{\rm 6.4~keV}$=1.6~keV, $\Gamma$=1.7), FE, 
and CXB are shown in red, blue, green, and cyan lines, respectively. The column densities of Abs1 and Abs2 are set to 
10$\times10^{22}$~H~cm$^{-2}$ and 6$\times10^{22}$~H~cm$^{-2}$, respectively.}
\label{gc_spec}
\end{figure}

\citet{ryu09} successfully applied this model to the X-ray spectra
of the 6.4~keV clumps near Sgr~B. 
In addition, \citet{muno04} and \citet{nobu10} reproduced the X-ray spectra of Sgr~A regions with similar models.
Therefore, we apply the same model to the Sgr~C region. 
In figure~\ref{gc_spec}, we show a simulated spectrum (XIS/FI; 0.5--10.0~keV) with the typical GC parameters (see caption). 
The plasma emission code used in this paper is APEC\footnote{Astrophysical Plasma Emission Code: A model of an emission spectrum from 
diffuse gas in collisionally-ionized equilibrium (\cite{smith01}).} and the absorption is evaluated in solar abundances\footnote{The solar abundance in this paper is referred to the values from \citet{anders89}.}.
 
\subsection{Spectral Fitting} 

\begin{figure*}[!ht]
  \begin{center}
   \FigureFile(135 mm,){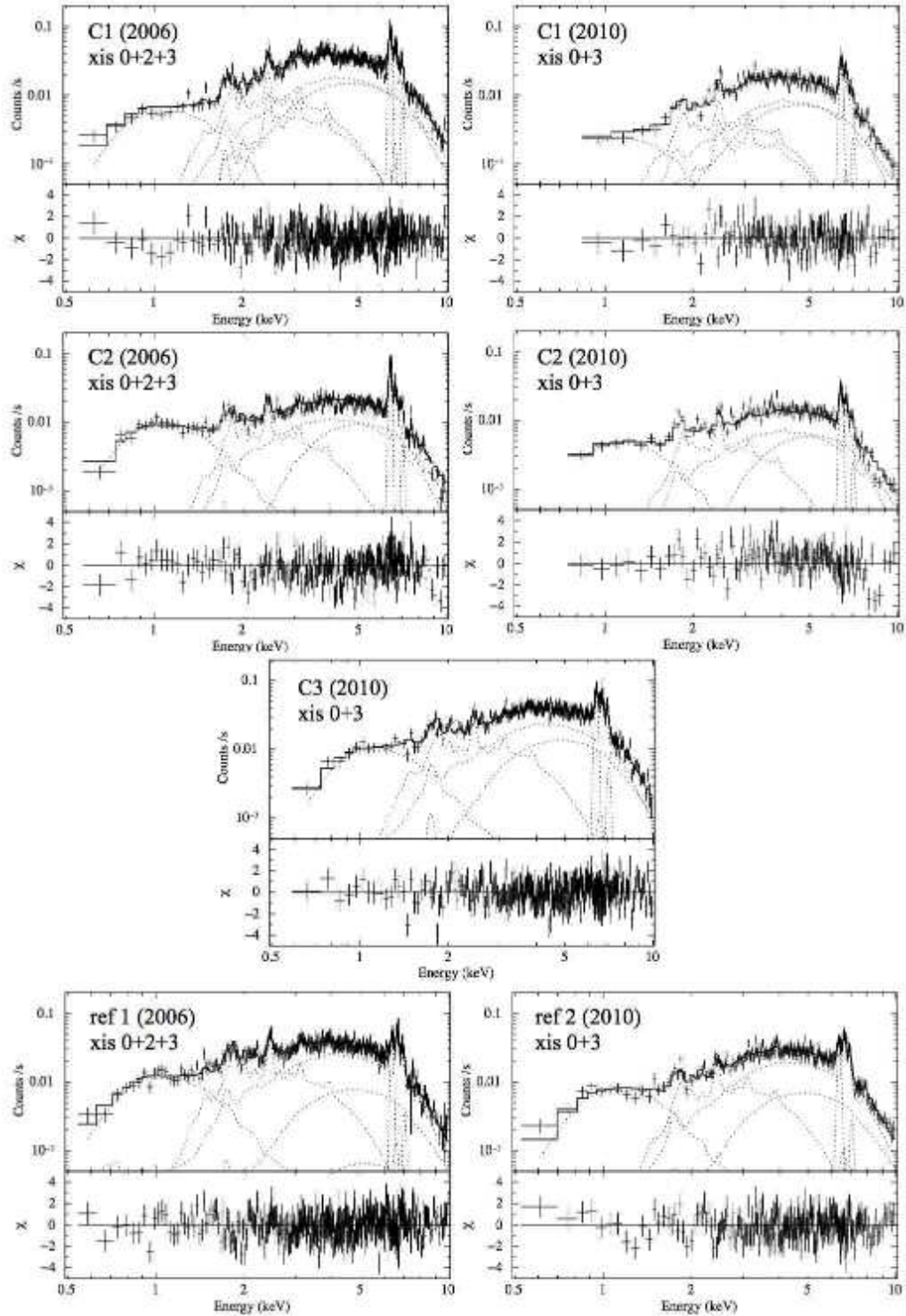}
   \end{center}
  \caption{Fitting results for all selected spectra (regions represented by black solid circles in figure~\ref{sgrc_img}a) near Sgr C. 
The FI (XIS~0+2+3 for 2006; XIS~0+3 for 2010) spectra with 1-$\sigma$ error bars and the best-fit models (c.f., figure~\ref{gc_spec}) are represented with solid crosses and dashed lines, respectively. 
The best-fit parameters are summarized in table~\ref{fit}.}
\label{all_spec}
\end{figure*}

As shown figure~\ref{all_spec}, we extract seven spectra from five regions (solid circles in figure~\ref{sgrc_img}).
The spectra of C1 and C2 are obtained twice in the observations of 2006 and 2010, whereas C3 is only observed in 2010.
All spectra are simultaneously fitted in the band of 0.5--10.0~keV with the model given in equation~(1).
Since the spectral model is highly complicated with many physical parameters, it is impractical to fit the data with all parameters free. 
Therefore, we adopt several reasonable constraints in the fitting. The parameter settings are the same as those of \citet{ryu09}.
The FE is fixed to an APEC model with k$T$=0.85~keV and abundance of $Z$=0.011, 
applied by small absorption of $N_{\rm H}$=0.17$\times10^{22}$~H~cm$^{-2}$ (\cite{ryu09}). 
The center energy of the Fe K$\alpha$ and K$\beta$ lines are respectively fixed to 6.4~keV and 7.05 keV with the flux ratio (K$\alpha$/K$\beta$) of 0.125 (\cite{kaa93}). 
The temperature of the HP (k$T_{\rm HP}$) is fixed to 6.5~keV (\cite{koyama07b}), and the abundances of HP and LP are fixed to one solar ($Z=1$). 
The parameter $\alpha$, normalization ratio of HP/LP, can be approximated to a constant near Sgr~C. This estimation is inferred from results of \citet{uchi12} (c.f., figure~3 and table~2 therein), which indicate that the two plasmas exhibit nearly the same spatial distribution with an e-folding length of $\sim$\timeform{0.6D}. 
\par

After all, the temperature of LP (k$T_{\rm LP}$), the normalization ratio of HP/LP ($\alpha$= norm$_{\rm HP}$/norm$_{\rm LP}$), the photon index ($\Gamma$) of the power law, and the equivalent width ($EW_{\rm 6.4~keV}$) of the 6.4 keV line are free parameters, but are linked in all regions including C1--3 and ref1--2. 
These settings are designed to verify a common and consistent XRN model.
Other parameters, such as line intensity ($I_{\rm 6.4~keV}$) and absorption ($N_{\rm H}$), are free and independent for each region.  
To investigate the time variability for C1 and C2,
the 6.4~keV intensities of the 2006 and 2010 spectra are set as free parameters, while other parameters are common between the two spectra.
\par
As is seen in figure~\ref{sgrc_img}b, C1 is entirely overlapped with the "Chimney",
which is a thermal plasma outflowing from G359.41$-$0.12. In order to include the Chimney contamination, we add the spectral model with fixed parameters (k$T$, $N_{\rm H}$, $Z$) reported in \citet{tsuru09}. 
The flux (normalization of Chimney) is calculated from the average surface brightness and the region size of C1 (radius=\timeform{2.0'}), which is 35\% of the GCPE flux.
\par
All seven spectra are generally reproduced by the model of equation~(1) with a reduced $\chi^{2}$/d.o.f of 3773/3031=1.26. 
The fitting results are shown in figure~\ref{all_spec} and the best-fit parameters are listed in table~\ref{fit}.
The large $EW_{\rm 6.4~keV}$ of $\sim1.2$~keV and the strong absorption ($N_{\rm H}$(Abs1)$\sim10^{23}$~H~cm$^{-2}$) agree with the XRN scenario for all the regions. 
The $N_{\rm H}$ of Abs2 are in the range of 5.0--6.5~$\times10^{22}$~H~cm$^{-2}$. 
These values are consistent with the interstellar medium (ISM) absorption toward the GC region (\cite{ryu09}; \cite{nobu10}; \cite{sakano02}).
As for the time variability  of the 6.4~keV line between the 2006 and 2010 observations, C1 exhibits a small increase of 8\% at 2.9-$\sigma$ level, while C2 shows no significant change. 
\par

For confirmation, we also apply independent fitting with free $EW_{\rm 6.4~keV}$ parameters for each region. The spectra of C1, C2, and C3 show 
$EW_{\rm 6.4~keV}$ of 1.1--1.5~keV, 1.1--1.6 keV, and 0.7--1.3 keV, respectively.
These values are consistent with the fitting result when $EW_{\rm 6.4~keV}$ is a common free parameter ($EW_{\rm 6.4~keV}$=1.15--1.27~keV; table~\ref{fit}), and agree with the XRN scenario of large $EW_{\rm 6.4~keV}$.
However, the independent $EW_{\rm 6.4~keV}$ fittings result in large statistical errors in the $R$ parameters (about 2--4 times larger than those in table~\ref{fit}), 
which is not practical for determining cloud positions (section~4.1) and following discussions.
Thus, we adopt the fitting results with $EW_{\rm 6.4~keV}$ as a common free parameter between the regions (table~\ref{fit}).

\begin{table*}[ht]
  \caption{Best-fit spectral parameters of the Sgr C regions$^{*}$. }
   \label{fit}
   \begin{center}
 \begin{tabular}{cccccccc}
 \hline
\hline     
     \multicolumn{8}{c}{Local parameters} \\ 
Region & Obs.Yr &$N_{\rm H}^\dagger$ (Abs1)&$N_{\rm H}^\dagger$ (Abs2)&$I_{6.4~\rm keV}$$^{\ddagger}$& FE norm$^\S$ &GCPE norm$_{\rm LP}$$^\|$&Fraction $R^\#$ \\
C1 & 2006  &6.5$^{+0.3}_{-0.5}$ &5.1$^{+0.3}_{-0.3}$ &1.60$^{+0.05}_{-0.05}$ & 0.060$^{+0.004}_{-0.004}$& 3.0$^{+0.1}_{-0.1}$& 0.23 $^{+0.03}_{-0.04}$ \\ 

   & 2010   &&& 1.73$^{+0.05}_{-0.05}$ &&&\\ 

C2 & 2006 &11.4$^{+1.6}_{-1.4}$ & 6.5$^{+0.3}_{-0.2}$ &1.51
$^{+0.03}_{-0.03}$  & 0.100$^{+0.005}_{-0.005}$ &2.0$^{+0.1}_{-0.1}$ & 0.72$^{+0.12}_{-0.03}$ \\ 

   & 2010  &&& 1.51$^{+0.03}_{-0.03} $&&&\\ 
C3 & 2010  &8.7$^{+0.4}_{-0.6}$ &5.0$^{+0.3}_{-0.2}$ &1.16$^{+0.05}_{-0.05}$   & 0.082$^{+0.005}_{-0.003}$ &3.0$^{+0.1}_{-0.1}$&0.26$^{+0.03}_{-0.03}$ \\ 

ref1 & 2006 &7.9$^{+0.5}_{-0.8}$ &5.3$^{+0.2}_{-0.1}$ &0.36$^{+0.03}_{-0.04}$ & 0.059$^{+0.003}_{-0.003}$ & 2.3$^{+0.1}_{-0.1}$& 0.40$^{+0.04}_{-0.03}$ \\ 

ref2 & 2010  &7.6$^{+0.5}_{-0.7}$  &6.4$^{+0.4}_{-0.3}$ &0.78$^{+0.05}_{-0.05}$ & 0.082$^{+0.006}_{-0.006}$ &3.3$^{+0.1}_{-0.1}$ & 0.31$^{+0.05}_{-0.05}$\\ 

\hline
\multicolumn{8}{c}{Global parameters} \\ 
\multicolumn{2}{c}{ k$T_{\rm LP}$ }&
\multicolumn{2}{c}{ $\alpha$ ratio ($\rm \frac{norm_{HP}}{norm_{LP}}$)  }&
\multicolumn{2}{c}{ $EW_{6.4~\rm keV}$ } &
Photon index $\Gamma$  \\ 
\multicolumn{2}{c}{ 0.92$^{+0.03}_{-0.05}$ keV }&
\multicolumn{2}{c}{ 0.28 $^{+0.01}_{-0.05}$}&
\multicolumn{2}{c}{ 1.19$^{+0.08}_{-0.04}$ keV } & 
1.63$^{+0.08}_{-0.09}$  \\ 

\hline

 \multicolumn{8}{c}{$\chi^{2}$/d.o.f\,$^{**}$ = $3773 / 3031 =1.26$}\\
\hline
\multicolumn{6}{@{}l@{}}{\hbox to 0pt{\parbox{165mm}
{\footnotesize
       \par\noindent
       \footnotemark[$*$] The spectral regions are shown in  figure~\ref{sgrc_img}. The uncertainties are at 90\% confidence level (statistical).
       \par\noindent
       \footnotemark[$\dagger$] The value of column density in the unit of $10^{22}$\,H\,cm$^{-2}$.
       \par\noindent
       \footnotemark[$\ddagger$] Intensity in the unit of $10^{-6}$ photons cm$^{-2}$s$^{-1}$arcmin$^{-2}$.
        \par\noindent
       \footnotemark[$\S$] The normalization factor of the APEC model for the foreground emission. It is normalized with the 
region size $A$ [arcmin$^2$] and expressed as $7.07 \times 10^{-13}/(4\pi D^2\,A)\,EM$ [cm$^{-5}$arcmin$^{-2}$], 
where $D$ and $EM$ are the distance to the source [cm], and the emission measure [cm$^{-3}$], respectively. 
       \par\noindent
  \footnotemark[$\|$] In the same expression as $\S$, but the normalization factor of LP in the GCPE.  
       \par\noindent
       \footnotemark[$\#$] $R$ is the fraction of the GCPE not absorbed by MC (see equation~(1) and figure~\ref{los}), 
which is the indicator of the line-of-sight position for XRNe (see equation~(2)). 
       \par\noindent
       \footnotemark[${**}$] The results are obtained by the simultaneous fitting of the 14 spectra from the FI and BI data.
   }\hss}
}
 \end{tabular}
  \end{center}
\end{table*}

\section{Results and Discussions}

\subsection{Line-of-sight Position of the XRNe } 
As illustrated in figure~\ref{los}, the parameter $R$ represents the line-of-sight position of the MC in the GCPE.  
Depending on the distance from Sgr~A*, the GCPE is more or less contaminated by the Galactic ridge plasma emission (GRPE), which has a similar two-temperature spectral structure but a more extended spatial distribution (\cite{uchi12}).  
To estimate the MC positions numerically, we use the spatial distributions of these plasmas.
\par
We introduce new coordinates ($X,Y$) with origins at Sgr~A* ($l=\timeform{-0.056D}, b=\timeform{-0.046D}$; \cite{yusef99}), in which $X$ and $Y$ correspond to the directions of $l$ and line of sight, respectively (see figure~\ref{los}). 
The offset of $b$ is ignored because the relevant XRNe are all on or near the Galactic plane (i.e., $|b|\leq$\timeform{0.1D}).
\citet{uchi11} and \citet{uchi12} derived the $X$-axis flux distribution for many emission lines of ionized atoms (e.g., Fe and S) in addition to the soft and hard X-ray continuum bands. 
The data were fitted with a two-exponential function along the Galactic plane ($X$-axis): 
$I_{\rm C}\exp(-X/l_{\rm C}) + I_{\rm R}\exp(-X/l_{\rm R}$). 
The values of e-folding length for both HP and LP are similar in the GCPE and GRPE.
In this paper, we adopted $l_{\rm C}$=\timeform{0.6D}, $l_{\rm R}$ = \timeform{50D}, 
and $I_{\rm C}/I_{\rm R}$=10 as the mean values (see table~2 in \cite{uchi12}).
 \par
 
We assume that the $X$-$Y$ plane distribution of the plasmas is an oblate spheroid
expanded with two-exponential function as
$I(X,Y)=A \times \exp(-r/r_{\rm GC}) + \exp(-r/r_{\rm GR})$, 
where $r=\sqrt{X^{2}+Y^{2}}$. 
$A$ is the flux ratio (GCPE/GRPE), and $r_{\rm GC}$ and $r_{\rm GR}$ are the e-folding scale lengths of GCPE and GRPE, respectively.  
Integrating $I(X,Y)$ along the $Y$-axis as $I(X)=\int^{\infty}_{-\infty} I(X,Y)
\;{\rm d}Y$, we find that the resultant projection of $X$-distribution is nearly 
the same exponential shape as observed by \citet{uchi12}.
We then compare $I(X)$ with the results of \citet{uchi12} to match the shapes. 
The parameters are nicely determined to be $r_{\rm GC}$=55~pc (or \timeform{0.4D}),  $r_{\rm GR}$=5000~pc (or \timeform{36D}), and $A$=900.

\par
\begin{figure}[!ht]
  \begin{center}
   \FigureFile(85 mm,){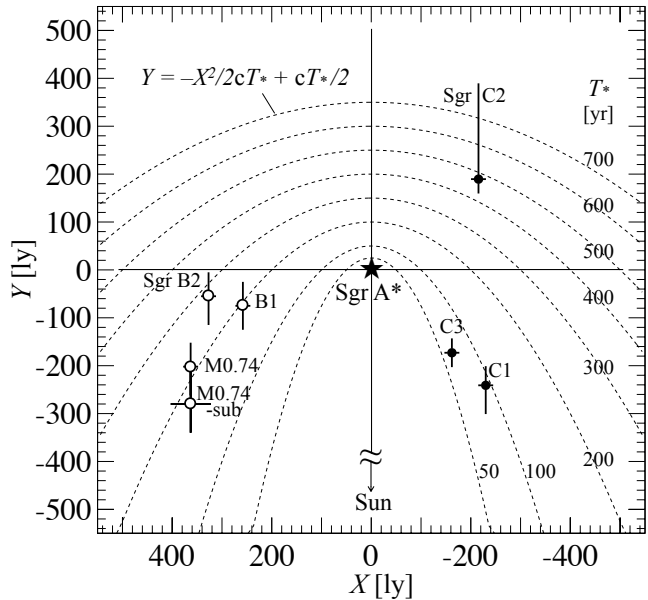}
   \end{center}
  \caption{Face-on view of Sgr~C XRNe (filled circles) around Sgr A*. The parabolas (dashed lines) represent the equal-time delay ($T_{*}$) contours for the X-ray echoes from Sgr A*. 
Sgr~B results from \citet{ryu09} (open circles) are also added (see section 4.3).
Parameter details are given in table~\ref{lumi}.}

\label{face_view}
\end{figure}

Considering that a MC/XRN is located at $(X_0, Y_0)$ and $X_0$ as its $l$ offset from Sgr~A*, the line-of-sight position ($Y_0$; also c.f., figure~\ref{los}) is determined from the following relation, using the obtained $R$ value (table~\ref{fit}).  

\begin{eqnarray}
\label{R}
 R(Y_0)=\frac{ \int^{Y_{0}}_{\rm -8\;kpc}  \; I(X_0, Y)\;\; {\rm d}Y }
             { \int^{\infty}_{\rm -8\;kpc} \; I(X_0, Y)\;\; {\rm d}Y }
\end{eqnarray}
We convert the unit of ($X, Y$) into the actual length of light year (ly). 
Figure~\ref{face_view} illustrates the resultant face-on view of XRNe around Sgr~A* on the $X$-$Y$ plane. 
Although C1 and C2 are proximately ($\sim$40~ly) located in the projection (figure~\ref{sgrc_img}a), C1 is in front of the $X$-axis, while C2 is largely behind C1 by $\sim$400~ly beyond the X-axis. 

\subsection{Association with the Radio Molecular Clouds} 
We have shown that C1, C2, and C3 are largely separated in the line-of-sight positions, although they are proximately located at the projected distance. 
We search for the radio counterparts of C1, C2 and C3 
from the CS ($J$=1$\rightarrow$0) maps (\cite{tsuboi99}) in different velocity ranges from $-$250~km~s$^{-1}$ to 
250~km~s$^{-1}$. As shown in figure~\ref{sgrc_img}c, C2 and C3 are respectively in good coincidence with MCs
in the velocity ranges of $-60\pm20$~km~s$^{-1}$ and +50$\pm30$~km~s$^{-1}$.
On the other hand, C1 is observed in the two separated velocity ranges of $-60\pm20$~km~s$^{-1}$ and $-100\pm20$~km~s$^{-1}$. 
Using the data corresponding C2 and C3, we derived a conversion factor between the X-ray column density (Abs1; table~\ref{fit}) and the integrated CS intensity (contours; figure~\ref{sgrc_img}c), 
which is [CS/$N_{\rm H}$]~$\simeq 2.5$~[K~km~s$^{-1}$/(10$^{22}$~H~cm$^{-2}$)].
The X-ray $N_{\rm H}$ of MC (Abs1) used in this study is determined directly from the absorption of GCPE continuum behind the cloud (c.f., figure~\ref{los}).
This method is available even if the MC is not an XRN, 
and thus does not depend on the properties of the illuminating source.
Assuming the  same [CS/$N_{\rm H}$] factor near Sgr~C, the observed X-ray column density of C1 
(6.0--6.8$\times$10$^{22}$~H~cm$^{-2}$; table~\ref{fit}) indicates a CS intensity of 15--17~K~km~s$^{-1}$. 
This generally excludes the $-$60~km~s$^{-1}$ MC as the counterpart of C1, because the average CS intensity is only 5--10~K~km~s$^{-1}$. 
The $-100$~km~s$^{-1}$ MC has a CS intensity of 10--25~K~km~s$^{-1}$, which is in agreement with the $N_{\rm H}$ estimated by the X-ray data. 
Thus, C1 is most likely to be situated in the velocity range of $-100\pm20$~km~s$^{-1}$.

\par
\citet{sofue95} identified two large $l$-$V$ structures known as "Arm~I" stretching from (\timeform{359.3D}, $-$150~km~s$^{-1}$) to (\timeform{0.9D}, +80~km~s$^{-1}$) and "Arm~II" 
stretching from (\timeform{359.4D}, $-$80~km~s$^{-1}$) to (\timeform{0.1D}, +60~km~s$^{-1}$). 
According to the velocities, the Sgr~C1 XRN may belong to Arm~I, while C2 is in Arm~II. 
Assuming the uniform circular rotation around Sgr~A*, \citet{sofue95} predicted that Arm~I and Arm~II are respectively located in the foreground and background with respect to Sgr~A* (c.f., figure 10 of \cite{sofue95}). 
The X-ray face-on view (figure~\ref{face_view}) is basically consistent with the radio result.

\subsection{XRN Parameters and Two-temperature Structure of the Galactic Center Plasma} 

The photon index and $EW_{\rm 6.4\;keV}$ for the 6.4~keV clumps in Sgr~C are 1.54--1.71 and 1.15--1.27 keV, respectively (table~\ref{fit}). These values are approximately equal to the Sgr~B results (\cite{ryu09}). 
The large $EW$ is in favor of the X-ray reflection and fluorescent origin (e.g., \cite{koyama96, mura01}). 
The photon index is consistent with the canonical active galactic nucleus (AGN; e.g., \cite{ishi96}), which favors the XRN scenario owing to the past flare of Sgr~A*.
\par 
 The best-fit temperature of LP (k$T_{\rm LP}$) and the mixing ratio  
 ($\alpha=\rm norm_{HP}/\rm norm_{LP}$) in the GCPE at Sgr~C 
are 0.87--0.95 keV, and 0.23--0.29, respectively. 
These values are nearly equal to those in Sgr~B region: k$T_{\rm LP}=$0.81--0.91~keV and $\alpha=$0.26--0.28 (\cite{ryu09}). 
Sgr~C and Sgr~B are located at nearly the symmetrical positions with 
respect to Sgr~A* on the Galactic plane, and hence these results indicate that the GCPE distribution is symmetrical not only in the flux profile (\cite{uchi12}) but also in the spectral shape. 
These results lead us to re-estimate the line-of-sight positions of Sgr~B MCs using the same methods for Sgr~C (section~4.1). 
The line-of-sight positions derived from parameter $R$ of the Sgr~B MCs (\cite{ryu09}) are also plotted in figure~\ref{face_view}. 

\subsection{Activity History of Sgr A* } 

On the basis of the XRN scenario of the past flares of Sgr~A*, 
we construct the parabolas, with their common foci at Sgr~A*, to illustrate 
the equi-delay time ($T_*$) contours of the X-ray echoes (see figure~\ref{face_view}).  
The time delay $T_{*}$ of the echo is given as below.
\begin{eqnarray}
\label{T}
T_{*} &=&(\sqrt{X^2+Y^2}+Y)/\rm c \hspace{10mm}[ yr] 
\end{eqnarray}
\par
\begin{figure}[!ht]
  \begin{center}
   \FigureFile(80 mm,){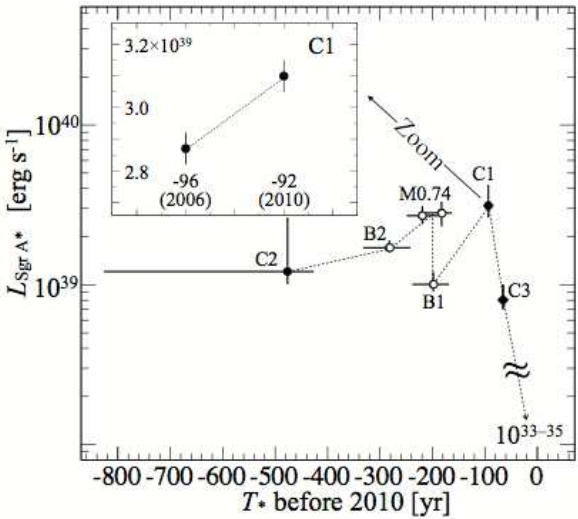}
   \end{center}
  \caption{
X-ray light curve of Sgr A* in the past 500 years. Data points of Sgr~C and Sgr~B are shown by filled and open circles, respectively. Error bars contain distance uncertainties (see table~\ref{lumi}). 
The inset panel shows the short-term variability (1-$\sigma$ error) indicated by 
the two observations of C1 on the 6.4~keV line (see table~\ref{fit}), in which the distance uncertainties are not included. The present luminosity of Sgr~A* is quoted from \citet{baga01} and \citet{porq03}.
}
\label{lc}
\end{figure}

Using the fluorescent Fe\emissiontype{I} intensity ($I_{6.4~\rm keV}$), the distance 
($D_{*}$=$\sqrt{X^2+Y^2}$), and the column density of XRN (Abs1; table~\ref{fit}), the required X-ray (2--10~keV) luminosity of Sgr~A* ($L_{*}$)  can be expressed as below (e.g., \cite{nobu08}; \cite{sun98}).

\begin{eqnarray}
\label{T}
 L_{*}&=& 1.0 \times (I_{6.4~\rm keV}~/ \rm  10^{-6}~photons~cm^{-2}s^{-1}arcmin^{-2}) \nonumber \\
      & & \times (D_{*}~/ \rm 300~ly)^2  \times  ({\it N}_{\rm H}~/ \rm 10^{22}~H~cm^{-2})^{-1} \nonumber \\
      & & \hspace{48mm} \rm [10^{40}~erg~s^{-1}]  
\end{eqnarray}
Here, the isotropic radiation of Sgr~A* with a photon index of $\Gamma=1.6$ is assumed (table~\ref{fit}). 
We then derive $T_{*}$ and $L_{*}$  for three Sgr~C XRNe (C1--3) and the Sgr~B XRNe (\cite{ryu09}). 
The results are summarized in table~\ref{lumi} and the long-term light curve of Sgr~A* is plotted in figure~\ref{lc}. 
Sgr~A* was active with short flares and possibly in a continuous high-luminosity state with $L_{*}=1$--$3\times10^{39}$ erg~s$^{-1}$ from the past $\sim$50 to $\sim$500 years ago. 
Thanks to the measurement of line-of-sight position of C2 (see figure~\ref{face_view}), we find that Sgr~A* was experiencing periods of high luminosity already 500 years ago,  
which extends the light curve of \citet{inui09} and \citet{ponti10} back by 200--400 years.


\begin{table*}[!ht]
  \caption{Three-dimensional parameters of XRNe and past luminosity of Sgr~A*. }
   \label{lumi}
   \begin{center}
 \begin{tabular}{llccclc}
 \hline
\hline     
XRN     &   Obs.  & $X^\dagger$ &  $Y^\dagger$ & $D_{*}^\S$ & $T_{*}^\|$ & $L_{*}^\#$ \\
 ID     &   Year  & [ly] & [ly] & [ly] & [year] &  [10$^{39}~$erg~s$^{-1}$] \\
\hline
 Sgr~C1 &   2010 &  -230 &  -241$^{+40}_{-60}$    & 333$^{+50}_{-30}$  & 92$^{+10}_{-10}$  &  3.1$^{+1.1}_{-0.5}$ \\
 Sgr~C2 &   2010 &  -216 &   189$^{+200}_{-30}$   & 287$^{+150}_{-20}$ & 476$^{+350}_{-50}$  & 1.2$^{+1.6}_{-0.2}$ \\%
 Sgr~C3 &   2010 &  -162 &  -173$^{+30}_{-30}$    & 237$^{+20}_{-20}$  & 64$^{+10}_{-10}$   &  0.8$^{+0.2}_{-0.1}$ \\
 Sgr~B2   &   2005  &  327 &  -55$^{+50}_{-60}$   & 331$^{+20}_{-5}$   & 281$^{+50}_{-40}$  &  1.7$^{+0.2}_{-0.1}$ \\
 Sgr~B1   &   2006  &  258 &  -75$^{+50}_{-50}$   & 269$^{+20}_{-10}$  & 198$^{+40}_{-30}$   &  1.0$^{+0.2}_{-0.1}$ \\
 M0.74$-$0.09 &2005 &  363 &  -202$^{+50}_{-60}$  & 416$^{+30}_{-20}$  & 219$^{+30}_{-30}$   &  2.7$^{+0.4}_{-0.3}$ \\
 M0.74-sub & 2007   &  363 &  -280$^{+80}_{-60}$  & 459$^{+40}_{-40}$  & 182$^{+40}_{-20}$   &  2.8$^{+0.5}_{-0.5}$ \\
\hline
\multicolumn{6}{@{}l@{}}{\hbox to 0pt
{\parbox{115mm}
	{\footnotesize
       \footnotemark[$*$] The uncertainties of $Y$ are at 90\% confidence level (statistical). 
       Errors of $D_{*}$, $T_{*}$, and $L_{*}$ are estimated using the $Y$ ranges.
       \par \noindent
       \footnotemark[$\dagger$] Coordinates $X$ and $Y$ correspond to the positions in the Galactic longitude ($l$) and the line of sight from Sgr~A*, respectively.  
       \par \noindent
       
       \footnotemark[$\S$] Distance from Sgr~A* calculated from $D_{*}$=$\sqrt{X^2+Y^2}$.       
       \par \noindent
       \footnotemark[$\|$] Time delay of echo from Sgr~A* calculated from equation~(3) and counted from 2010.

       \footnotemark[$\#$] Required past X-ray luminosity of Sgr~A* calculated from equation~(4), also see text.           
 	} 
      ã\hss}
} 
\end{tabular}
  \end{center}
\end{table*}

In the past high state, the luminosity shows sporadic variabilities (short flares) 
at $\sim$200 and $\sim$100 years ago as suggested by Sgr~B (M~0.74) and Sgr~C1, respectively.
The short flare at 200--300 years ago would be responsible for the short-term time variability of Sgr~B2 (\cite{nobu11}; \cite{koyama08}), and hence the variability time scale would be 5--10 years. 
The other short flare at $\sim$100 years ago has similar time-scale as inferred by the 6.4~keV time variability (increase) of C1 in 4 years (2006--2010) at 3-$\sigma$ level (table~\ref{fit}). 
This flare may correspond to the time variability of the Sgr~A XRNe. 
\citet{muno07} and \citet{koyama09} reported variability of the 6.4 keV line from the Sgr~A cloud in a short time-scale of 3--5 years at a $\sim5\sigma$ sigma level.
Then \citet{ponti10} and \citet{cap12} predicted the flare-like events occurred at $\sim$100--400 years ago, which is basically consistent with our results.	
\citet{ponti10} proposed that the flare should have started $\sim$400 years ago, while \citet{cap12} suggested that a high period ended $\sim$150 years ago from the illumination pattern in the Sgr A complex. 
However, the estimated flare luminosities are different: 
$L_{*} \sim 10^{39}$ erg~s$^{-1}$ was suggested by \citet{ponti10}, while $L_{*}\sim10^{38}$ erg~s$^{-1}$ was suggested by \citet{cap12}.
These inconsistencies are mainly attributable to unclear
line-of-sight positions and partly due to the difficult estimation of $N_{\rm H}$ for the Sgr~A XRNe. 
The luminosity determined by \citet{cap12} is $\sim10^{37}$--$10^{38}$ erg~s$^{-1}$ 
around 70 and 130 years ago, which is about 10 times lower than that obtained here: $L_{*}\sim10^{39}$ erg~s$^{-1}$ at $\sim$90 years ago. 
The high luminosity indicated by C1 may be  attributed to a short-term flare that occurred in the fading period reported by \citet{cap12}.
A unified analysis for the Sgr~A XRNe as performed in Sgr~C and Sgr~B in this paper would slove these problems.
\par

The average X-ray luminosity in the high state  was $L_{*}^{\rm ave}\simeq2\times$10$^{39}$~erg~s$^{-1}$,
which is only $\sim4\times10^{-6}$ of the Sgr~A* Eddington limit 
($L_{\rm E}\simeq 5\times$10$^{44}$~erg~s$^{-1}$) with a mass of 4$\times10^{6}$~M$_{\solar}$ (\cite{ghez08}).
Still, from this luminosity, we may state that Sgr~A* has been a low-luminosity AGN (e.g., M~81; \cite{ishi96}) for the past 500 years.
\par  
The present X-ray luminosity of Sgr~A* is $L_{*}\sim10^{33-35}$~erg~s$^{-1}$ (\cite{baga01}; \cite{porq03}),
and hence  a sudden luminosity drop by 4--6 orders of magnitude should have occurred within $\sim$100 years.
\citet{sch10} reported that the quasar IC~2497 with mass of $\sim10^{9}$~M$_{\solar}$ experienced an dramatic luminosity drop by more than 4 orders of magnitude in 45000--70000 years. 
On the other hand, for the Galactic X-ray binaries with a black hole mass of $\sim$10 M$_{\solar}$ (e.g., GRS 1915+10; \cite{morgan96}), a state change from high/soft to low/hard was detected on a time scale of 1~hour. 
Assuming that the state-change time scale is proportional to black-hole mass, 
at a mass of 4$\times10^{6}~\rm M_{\solar}$,
70000 years (for $10^{9}~\rm M_{\solar}$) and 1 hour (for $10~\rm M_{\solar}$) correspond to 280 years and 46 years, respectively.
Then, the time scale of the luminosity drop for Sgr~A* is estimated to  
be as fast as 50--300 years, which is consistent with the prediction in our light curve (figure~\ref{lc}).

\vspace{5mm}
\newpage

We are grateful to all members of the Suzaku hardware and software teams. This work is supported by the Grant-in-Aid for the Global COE Program ``The Next Generation of Physics, Spun from Universality and Emergence'' from the Ministry of Education, Culture, Sports, Science and Technology (MEXT) of Japan.
SGR and SM are financially supported by the Japan Society for the Promotion of Science (JSPS). TGT is supported by JSPS Scientific Research~B Nos. 20340043 and 23340047. 
KK is supported by JSPS KAKENHI Grand Numbers 23000004 and 24540229.
MN is also supported by JSPS KAKENHI Grand Number 24740123.

\end{document}